\documentclass{emulateapj}

\def\be{\begin{eqnarray}}
\def\ee{\end{eqnarray}}
\def\etal{et al.}

\def\h{_{\rm H}}
\def\hi{_{\rm HI}}
\def\hii{_{\rm HII}}
\def\hiiQ{_{\rm HII,Q}}
\def\hiiobs{_{\rm HII,obs}}
\def\hiimod{_{\rm HII,mod}}
\def\bias{^{\rm bias}}

\def\PS{_{\rm PS}}
\def\PShalo{_{\rm PS,halo}}
\def\ST{_{\rm ST}}
\def\SThalo{_{\rm ST,halo}}
\def\Q{_{\rm Q}}
\def\halo{_{\rm halo}}
\def\star{_{\rm star}}
\def\min{_{\rm min}}

\def\phsQ{_{\rm phs,Q}}
\def\phsstar{_{\rm phs,star}}
\def\rec{_{\rm rec}}
\def\duty{_{\rm duty}}
\def\HIGM{_{\rm H,IGM}}
\def\vir{_{\rm vir}}
\def\rQ{_{r_{\rm Q}}}
\def\crit{_{\rm crit}}

\def\eV{{\rm\,eV}}
\def\kms{{\rm\,km\,s^{-1}}}
\def\msun{{\rm\,M_\odot}}
\def\Mpc{{\rm\,Mpc}}
\def\kpc{{\rm\,kpc}}
\def\yr{{\rm\,yr}}
\def\Gyr{{\rm\,Gyr}}
\def\cm{{\rm\,cm}}
\def\ps{{\rm\,s}^{-1}}

\begin{document}
\title{The Str\"omgren sphere, the environment and the reionization
in the local universe of the highest redshift QSOs
}
\author{Qingjuan Yu\footnotemark[1]}
\affil{Department of Astronomy, 601 Campbell Hall, University of California
at Berkeley, Berkeley, CA 94720}
\author{Youjun Lu}
\affil{Department of Astronomy, 601 Campbell Hall, University of California
at Berkeley, Berkeley, CA 94720;\\Center for Astrophysics,
University of Science and Technology of China, 96 Jinzhai Road, Hefei,
Anhui 230026, P.\ R.\ China
}
\footnotetext[1]{Hubble Fellow}
\email{yqj,lyj@astro.berkeley.edu}

\begin{abstract} In this paper we investigate the environment and reionization
process around the highest redshift QSOs having Gunn-Peterson troughs
($z>6.1$).  Starting with the cosmic density perturbation and structure
formation theory and the fact that the highest redshift QSOs are located in
rare overdense regions, we show that the halo formation, gas distribution,
and star formation around QSOs are biased from those of the cosmic
average.  We argue that a significant fraction of hydrogen in the
Str\"omgren sphere around QSOs is ionized by photons from stars and that only
about several percent to at most 10\%--20\% of the total hydrogen is left
(e.g., in minihalos, halos, or high-density subregions) to be
ionized by QSO photons.
The cosmic average neutral hydrogen fraction at $z\sim 6.2-6.4$ should
also be smaller than the upper limit of 10\%--20\% and may be only
a few percent.
We analyze the clumping property of the hydrogen ionized by
QSOs and study the evolution of the Str\"omgren sphere.  We find that the
expected Str\"omgren radii from our models are consistent with observations if
the lifetime of the highest redshift QSOs is about or longer than a few times
$10^7\yr$ (as is the lifetime of the main population of QSOs; with comoving number density peaked
at $z\sim$2--3). With such a QSO lifetime, the ages of most of the observed QSOs
are long enough that the QSO photon emission is balanced by the recombination
of the hydrogen ionized by QSO photons in their Str\"omgren spheres,
and the expected Str\"omgren radii from the balance are independent
of the detailed values of the QSO ages.  We also point out a statistical method
involving a larger sample of QSOs having Gunn-Peterson troughs in future
observations, which may potentially check or rule out the possibility that the
highest redshift QSOs have a shorter lifetime (e.g., $<10^7\yr$), even without
an accurate estimate of the hydrogen clumping property.

\end{abstract}
 \maketitle

\section{Introduction}\label{sec:intro}

QSOs at redshifts $z>6.1$ have been discovered by the Sloan Digital Sky Survey
(SDSS) recently (\citealt{Fanetal01,Fanetal03,Fanetal04}).  Their spectra have
shown the existence of the Gunn-Peterson absorption trough \citep{GP65,S65} at
wavelengths blueward of the Ly$\alpha$ emission line, which suggests that
hydrogen in the early universe is significantly neutral ($\ga 1\%$ in mass
average; \citealt{Fanetal01,Fanetal02,Becker01,D01}).
Despite the presence of the Gunn-Peterson trough in the spectra,
a significant amount of flux is transmitted at wavelengths blueward of but
near the center of the Ly$\alpha$ line, which suggests that the regions
surrounding these highest redshift QSOs are highly ionized by QSO photons
(i.e., the proximity effect of QSOs, e.g., in \citealt{BDO88}). These
highly ionized regions are generally idealized as spherical and called
``Str\"omgren spheres.'' Their radii (or the Str\"omgren radii) can be
estimated from the wavelength difference between the Gunn-Peterson trough and
the Ly$\alpha$ line center (e.g.,\citealt{White03}).

Physically the sizes of the highly ionized regions
are closely related to not only the neutral hydrogen distribution in the local
universe of these QSOs before their nuclear activities turn on but also the
subsequent QSO luminosity evolution, either of which thus may be constrained by
the Str\"omgren radii estimated from QSO spectra provided that the range of the
other one is given.  For example, using the observationally determined
Str\"omgren radii and assuming the neutral hydrogen fraction (defined by the
ratio of the neutral hydrogen number density to the total hydrogen number density)
$x\hi=1$, \citet{HC02} found that the lifetime of one highest redshift QSO
is not shorter than $2\times 10^7\yr$ (see also Pentericci et al.\ 2002) and \citet{White03} got a substantially
shorter lifetime for two;
using the measured Str\"omgren radii of two highest redshift QSOs and some
independent constraints on the QSO lifetime obtained by other studies,
\citet{WL04} argued that $x\hi$ may be higher than tens of percent at $z\sim
6.3$, which is significantly higher than the lower limit of $x\hi \ga 1\%$
constrained by the flux upper limit of the Gunn-Peterson absorption trough
\citep{Becker01,Fanetal02, White03}.  In addition, \citet{CH00} and
\citet{MR00} investigated the structure of the transmitted spectra at
wavelengths blueward of the Ly$\alpha$ line and pointed out that the red
damping wing of the Gunn-Peterson trough might leave detectable features there
\citep{Miralda98}, which may also help to constrain $x\hi$ (see a recent
study by \citealt{MHC04} using mock data from simulations.

In most previous investigations on the reionization process around the highest
redshift QSOs, the QSOs are implicitly assumed to be located in cosmic average
regions. However, in the hierarchical formation and evolution scenario of
galaxies and QSOs \citep[e.g., ][]{KH00}, luminous QSOs (especially those at
high redshifts $z>6.1$) are located in rare overdense regions. The structure
formation in these regions and consequently the gas distribution, star
formation, and reionization of the neutral hydrogen may be significantly
biased away from those in the regions with cosmic mean density.
Some discussions in a few recent papers are also related to some biasing
effects of a dense environment
(e.g., \citealt{Ciardi03,Cen03b,GP04,WL04b,Furlanetto04,Santos}).
The purpose of
this paper is to quantitatively investigate the dense environment of the
highest redshift QSOs, the reionization in their local universe due to ionizing
photons from both stars and QSOs, and the
evolution of their Str\"omgren spheres.  Comparison of the expected Str\"omgren
radii from the more realistic model with observations would improve our
understanding of the universe's reionization history and the formation and
evolution of QSOs.

This paper is organized as follows. The observational properties of the highest
redshift QSOs that have the Gunn-Peterson trough are reviewed in
\S~\ref{sec:data}. In \S~\ref{sec:environ}, we investigate the environment
around these QSOs in the framework of the cosmic density perturbation and
structure formation theory \citep[e.g.,][]{LC93}.  We first estimate how dense
the environment is from the number density of the highest redshift QSOs.
Then we introduce a hybrid model \citep[see also][]{BL04} based on
the excursion set approach of the Press-Schechter formula
\citep{Bondetal91,LC93} to see how the structure or halo formation is enhanced
in these rare overdense regions in \S~\ref{sec:subhaloes}. Gas in halos may be
shock heated, collisionally ionized, and then cool to form stars, galaxies,
and/or QSOs.  Ionizing photons from stars and QSOs will jointly contribute
to the reionization of the surrounding neutral hydrogen.
In this paper we isolate their effects and first consider the reionization due
to stars (the variable $x\hi$ in this paper represents the neutral hydrogen
fraction obtained after considering the reionization due to stars but before
the reionization due to the QSO becomes effective). The justification for this isolation
is that before the cosmic time that the QSO redshift corresponds to,
the QSO ionizing photon emission rate was high enough to be comparable
to (or higher than) the ionizing photon emission rate from stars only in a
short period (a few
times $10^7\yr$) and the cumulative number of ionizing photons are mainly
from stars (see detailed discussion in \S~\ref{sec:subreion} and
\ref{sec:model}).
In \S~\ref{sec:subreion}, using some
simple star formation models, we find that more than 80\%--90\% of neutral
hydrogen in these dense regions is ionized by photons from stars. Inhomogeneity
or clumpiness of hydrogen may significantly affect the reionization process.
In \S~\ref{sec:clump} we give a rough estimate of the clumpiness of
hydrogen ionized by QSO photons, based on the halo distribution obtained in
\S~\ref{sec:subhaloes}, some numerical simulation results obtained in other
studies, and some physical arguments.  In the environment of the highest
redshift QSOs analyzed in \S~\ref{sec:environ}, we study the evolution of the
apparent size of the Str\"omgren sphere highly ionized by QSO photons in
\S~\ref{sec:Rsph}.  The model results are compared with observations and the
implications are discussed.  The main conclusions are summarized in
\S~\ref{sec:con}.

Throughout this paper, we adopt a standard $\Lambda$CDM cosmological model with
$\Omega_{\Lambda}=0.73$, $\Omega_{\rm m}=0.27$, $\Omega_{\rm b}=0.0444$,
$H_0=71\kms\Mpc^{-1}$, $\sigma_8=0.84$, and $n=0.93$ \citep{Spergel03}.

\section{Sample of the highest redshift QSOs that have Gunn-Peterson 
troughs} \label{sec:data}

To date, four QSOs have been observed at redshift $z>6.1$, and all of them
are detected to have the Gunn-Peterson trough in their spectra
\citep{White03,Fanetal03,Fanetal04} and are compiled in Table~\ref{tab:tab1}.  As
seen from Table~\ref{tab:tab1}, the redshifts of the observed QSOs are in the
range of $z\Q\simeq$6.2--6.4.  The redshift may be determined
from the Ly$\alpha$ line (e.g., for SDSS J1623+3112; \citealt{Fanetal04}), and
the typical redshift error is $\pm0.02$, which is mainly caused by
the uncertainty in the determination of the location of the line center in
the spectra.  The redshift of SDSS J1148+5251 is precisely determined
from CO and Mg$_{\rm II}$ lines \citep{Walter03,WMJ03}, and we simply assign a small
error of $\pm 0.005$ to its redshift (similarly for SDSS J1048+4637 whose
redshift is determined by Mg$_{II}$ line; \citealt{Maiolino}). Although the redshift
of SDSS J1030+0524 has also been determined from metal lines (e.g., C$_{\rm IV}$), 
the redshift error can still be as large as 0.02, caused by the possible
blueshift of the C$_{\rm IV}$ line \citep[e.g.,][]{Richards02}, and we therefore set 
its redshift error to be $+0.02$ and $-0.005$.
The Str\"omgren radii of three QSOs in Table~\ref{tab:tab1} have been
estimated through the difference between the QSO redshift and the redshift of
the onset of the Ly$\alpha$ Gunn-Peterson trough in their 
spectra in the literature (i.e., $r\hiiobs^0$; see also
$z\hiiobs$ in Tab.~\ref{tab:tab1} for the corresponding redshift range of the
Str\"omgren sphere, except for SDSS J1048+4637). The errors on $r\hiiobs^0$
in Table~\ref{tab:tab1} are introduced by the uncertainty
in the determination of these QSO redshifts.  

There may also exist uncertainties in the determination of the onset 
redshift of the Gunn-Peterson troughs, which further introduce uncertainties
in the estimate of the Str\"omgren radii. For example,
\citet{MH04} point out that the radius estimated through the Ly$\alpha$ trough may
only give a lower limit of
the Str\"omgren radius and that a better estimate may be obtained by using
the Ly$\beta$
trough.  Using the Ly$\beta$ trough and some numerical simulation results,
\citet{MH04} increase the estimate of the Str\"omgren radius of SDSS J1030+0524
to $\sim 6\Mpc$ (from $\sim 4.6\Mpc$ listed in Tab.~1).  Note that the new
estimate might be affected by the numerical simulation used in the analysis
\citep{MH04}, which does not well represent the overdense environment of the
highest redshift QSOs discussed in this paper. In addition, other QSOs listed
in Table~\ref{tab:tab1} do not have the Str\"omgren radius estimated by using
the Ly$\beta$ trough in the literature, and thus we still mainly use the values
estimated from the Ly$\alpha$ troughs for consistency in this paper and discuss
the implication of a possible systematically larger value of the observational
Str\"omgren radius in \S~\ref{sec:discussion}.

We estimate the ionizing photon emission rates of these QSOs, $\dot{N}^0\phsQ$,
by using their absolute magnitude $M_{1450}$ at $1450~\AA~$ listed in Table~\ref{tab:tab1}
(compiled from \citealt{Fanetal01,Fanetal03,Fanetal04}) and the 
spectral templates of relatively low redshift QSOs ($z\la 3$) shown
in \citet{Telfer02}.
(\citealt{Fanetal04} have found that the average spectrum of the highest redshift
QSOs is similar to that in \citealt{Telfer02}.)
In \citet{Telfer02}, the average spectra of the low-redshift 
QSOs (energy radiated per unit time and per logarithmic interval of
frequency $\nu f_\nu\propto\nu^{1+\alpha}$) is well fitted by a double power
law in the optical to EUV band.  For
radio-loud QSOs, the EUV and near-UV spectral indices are $\alpha_{\rm
EUV}=-1.96\pm0.12$ and $\alpha_{\rm NUV}=-0.67\pm0.08$, respectively,
with the break at $\sim 1280\AA$, while for radio-quiet
QSOs they are $\alpha_{\rm EUV}=-1.57\pm0.17$ and
$\alpha_{\rm NUV}=-0.72\pm0.09$, respectively. 
We count photons with energy in the range 13.6--54.4$\eV$
as ionizing photons.
Since so far little information on the radio properties of the
highest redshift QSOs is available, we obtain two emission rates for each QSO
listed in Table~\ref{tab:tab1}, the smaller of which is obtained 
by adopting the spectral indices of radio-loud QSOs and the larger of which is
obtained by adopting the spectral indices of radio-quiet QSOs.
We take the middle point of the whole range of the two values
(including their errors) as $\dot{N}^0\phsQ$ and take the whole range as the
error range of $\dot{N}^0\phsQ$ (see also the filled circles and their
horizontal error bars in Fig.~\ref{fig:f3}a).
Note that for SDSS J1030+0524, compared to the rate listed in
Table~\ref{tab:tab1}, a substantially smaller rate is obtained in \citet{MH04}
by fitting the spectrum of the QSO in a surrounding hydrogen density
distribution given by a numerical simulation.  As mentioned above, 
the numerical simulation used in \citet{MH04} mimics the cosmic average
environment rather than the rare overdense environment around
the highest
redshift QSOs; thus the smaller rate may be an underestimate since more
ionizing photons would be required to balance the recombination in a
denser environment.

The highest redshift QSOs are extremely rare and their comoving number density
$N_{\rm Q}$ is roughly the same as the density of the QSOs with absolute
magnitude $M_{1450}<-26.7$ at $z\sim 6$ estimated by
\citet{Fanetal03,Fanetal04}, i.e., $N_{\rm Q} \simeq (6\pm 2) \times
10^{-10}\Mpc^{-3}$.  (Note that the value of $N_{\rm Q}$ is obtained by
assuming a cosmological model with $H_0=65\kms\Mpc^{-1}$, $\Omega_{\rm m}=0.35$,
and $\Omega_{\Lambda}=0.65$; we have adjusted this value to the cosmological
model adopted in this paper and found no significant difference.)  

\section{The environment around the highest redshift QSOs } \label{sec:environ}

The highest redshift QSOs are located in rare overdense regions according to
current structure and QSO formation and evolution picture in the $\Lambda$CDM
cosmogony \citep[e.g.,][]{KH00}. To quantify the environment of these QSOs, we
first estimate the mean mass overdensity in their surrounding regions in
\S~\ref{sec:overden}. Then we study the structure and halo formation in
these overdense regions in \S~\ref{sec:subhaloes}, and study
the star formation and the reionization due to ionizing photons emitted from
these stars in \S~\ref{sec:subreion}. In \S~\ref{sec:clump} we study
the clumpiness of hydrogen in the Str\"omgren sphere of the QSOs.

\subsection{Overdensity}\label{sec:overden}

The mean overdensity within a sphere with proper radius $r$ at redshift $z$
is defined
by $\bar\delta_{r}\equiv[\langle \rho \rangle_{r}-\bar\rho]/\bar\rho$,
where $\langle \rho \rangle_{r}$ is the average comoving mass density in the
sphere and $\bar\rho$ is the comoving cosmic mean mass density. 
We denote the comoving radius of the sphere by $R=r(1+z)$. (Here we do not
distinguish the difference between the Eulerian radius and the Lagrangian
radius, since the overdensity considered in this paper is small.)
The variance of the density fluctuations at redshift $z$ in spheres with
comoving radius $R$, $\sigma^2(R,z)$, can be determined by \citep[e.g., see
eq.~2.55 in][]{SWhite}
\be
\sigma^2(R,z)=D^2(z)\int^{\infty}_0 \frac{dk}{2\pi^2}k^2 P(k) W^2(kR),
\label{eq:sigmz}
\ee
where $D(z)$ is the linear growth factor of perturbations at $z$ and normalized
to 1 at the present time, $P(k)$ is the power spectrum of the density
perturbation field and is computed with the fitting formula of
\citet{EH99}, and $W(kR)=3[\sin(kR)-kR\cos(kR)]/(kR)^3$ is the top-hat
window function.
If the perturbations on the scale of $r$
are still in the linear growth regime at redshift $z$, the mean
overdensity $\bar\delta_r$ follows a Gaussian distribution:
\be
&&p(\bar\delta_r,z)d\bar\delta_{r} \nonumber \\
&&=\frac{1}{\sqrt{2\pi}\sigma(R,z)}
\exp\left[-\frac{\bar{\delta}^2_{r}}{2\sigma^2(R,z)}\right]
d\bar\delta_{r},
\label{eq:denprob}
\ee
and the probability that a random region with scale $r$ has a mean
overdensity higher than a value $\bar\delta_r'$ is given by
\be
{\cal P}(>\bar\delta_r',z)=\int^{\infty}_{
\bar\delta_r'}p(\bar{\delta}_{r},z)d\bar{\delta}_{r}.
\label{eq:overprob}
\ee

Since the Str\"omgren radii measured from the observed QSO spectra are around
$5\Mpc$, below we consider a sphere with proper radius
$r\Q=5\Mpc$.  According to equation
(\ref{eq:sigmz}), the standard deviation of the density fluctuations in this
sphere at $z\simeq 6$ is $\sigma[R=r\Q(1+z), z\simeq 6]\simeq 0.06$, which is
much smaller than the critical overdensity $\delta_c=1.68$ for dynamical
collapse, and the perturbations are still in the linear growth regime. Thus
equations (\ref{eq:denprob}) and (\ref{eq:overprob}) can be applied to these
spheres.

In regions with sufficiently high overdensity, sufficiently massive
halos will form and these halos are the hosts of luminous QSOs.
In a random spherical region with proper radius $r\Q=5\Mpc$ at $z\sim 6$,
the probability of finding a halo capable of hosting nuclear-active QSOs with
$M_{1450}<-26.7$ (e.g., with mass $\sim10^{13}\msun$) is
\be
f^{-1}\duty N_{\rm Q}\frac{4\pi}{3}[r\Q(1+z)]^3\simeq 1.0\times 10^{-4}f^{-1}\duty,
\label{eq:Pduty}
\ee
where $f\duty(z)$ represents the duty cycle of QSOs, defined to be the number
ratio of the nuclear active QSOs to the halos capable of hosting the QSOs at
redshift $z$.  The halos capable of hosting QSOs include both those
hosting a nuclear-active QSO and those hosting a dead QSO with nuclear activity
quenched.  For the highest redshift QSOs, we simply assume $f\duty\sim 1$
because most sufficiently massive halos (e.g., $\ga10^{13}\msun$) at $z\sim 6$
are actually formed within a period (from $z=7$ to $6$; e.g., see Fig.~1 in
\citealt{MW02}) not significantly longer than the QSO lifetime ($>$ a few
times $10^7\yr$; \citealt{YT02,YL04a}; see more discussions about the QSO
lifetime 
in \S~\ref{sec:Rsph}).  By multiplying equation (\ref{eq:overprob}) by
a factor of 2 (this factor comes from the so-called cloud-in-cloud
problem; see discussions in \citealt{Bondetal91}) and then setting it equal to 
equation (\ref{eq:Pduty}), we obtain that the mean overdensity
of the regions associated with the highest redshift QSOs is
$\bar{\delta}\rQ\ge\bar\delta\rQ'\sim 4 \sigma(R,z\simeq 6)\simeq 0.25$.
Since the overdensity distribution
$g(\bar{\delta}_{r})$ decreases exponentially at the high overdensity end, most
of these overdense regions associated with the highest redshift QSOs do not 
have a mean overdensity significantly higher than $\bar\delta\rQ'$,
but near the value.  

Note that the connection of equation (\ref{eq:overprob}) with equation
(\ref{eq:Pduty}) above has assumed a one-to-two correspondence between
the likelihood of finding a region with mean overdensity above a certain
value and the likelihood of a same-sized region containing a halo capable of
hosting a quasar. Although in principle there might exist regions with
overdensity above the certain value but not containing such a halo,
the simple approach above is good enough in practice
for the rare overdense regions and for the purpose of this paper.
For example, given a halo with a certain mass, 
the method in Barkana (2004; for simplicity 
the details are not presented here), which is also based on the (extended)
Press-Schechter
formalism, may provide the expected average overdensity within a certain region
around the halo and the standard deviation of the overdensity.
Assuming that the host halos of the highest redshift QSOs are around
$10^{13}\msun$ (or $2\times 10^{13}\msun$) and using the method in \citet{Barkana04}, we obtain an
average overdensity of $\sim 0.13$ (or $0.16$)
within the region with radius $r\Q$ surrounding the halo, which is smaller
than the estimate above.
Hereafter, we simply adopt
$\bar{\delta}\rQ=\bar\delta\rQ'\simeq 0.25$ as the mean overdensity of the
spheres with proper radius $r\Q=5\Mpc$ centering on the highest redshift QSOs.
Using the value of $\bar{\delta}\rQ=0.13$ will not qualitatively affect the
conclusions in this paper.

\subsection{The formed structure}\label{sec:subhaloes}

In this section, we introduce a hybrid model to estimate the mass distribution
of halos formed in the rare overdense regions associated with the highest
redshift QSOs (see also \citealt{BL04}).

We define the halo mass function as $\frac{dN}{dM}(z)$ so that
$\frac{dN}{dM}dM$ represents the comoving
number density of halos with mass in the range $M\rightarrow M+dM$ at redshift $z$.
In the cosmic average regions, the halo mass function can be given by
\be
\frac{dN}{dM}=\frac{\bar{\rho}}{M}\left|\frac{d\sigma^2(R_c,z)}{dM}
\right| f[\delta_c,\sigma^2(R_c,z)],
\label{eq:mf}
\ee
where  $R_c=(3M/4\pi\bar{\rho})^{1/3}$, and
$f[\delta_c,\sigma^2(R_c,z)]d\sigma^2(R_c,z)$ is the mass
fraction of halos with mass in the range $M\rightarrow M+dM$ at redshift $z$. 
In the model of \citet{PS74}, $f[\delta_c,\sigma^2(R_c,z)]$
is given by
\be
f\PS[\delta_c,\sigma^2(R_c,z)]=\frac{1}{\sqrt{2\pi}}\frac{\nu}
{\sigma^2(R_c,z)}\exp(-\frac{\nu^2}{2}),
\label{eq:mfPS}
\ee
where $\nu=\delta_c/\sigma(R_c,z)$. Compared to the results obtained from
cosmological simulations with very large volumes, however, the Press-Schechter
mass function substantially underestimates the abundance of the rare massive
halos that host galaxies at high redshifts and overestimates the abundance of
intermediate-mass halos. To more accurately match the simulation results,
\citet{ST99} introduce a new formula
\be
&& f\ST[\delta_c,\sigma^2(R_c,z)] \nonumber \\
&& =\frac{A{\arcmin}}{\sqrt{2\pi}} \frac{\sqrt{a{\arcmin}}\nu}
{\sigma^2(R_c,z)} \left[ 1+\frac{1}{({a{\arcmin}}\nu^2)^{q{\arcmin}}}\right]
\exp\left(-\frac{{a{\arcmin}}\nu^2}{2}\right)
\label{eq:mfST}
\ee
with the best-fit parameters $a{\arcmin}=0.707$ and $q{\arcmin}=0.3$, and
the normalization parameter $A{\arcmin}=0.322$. This formula can be derived
by introducing an ellipsoidal collapse model instead of the usually adopted
spherical collapse model into the excursion set approach of the Press-Schechter
formula \citep{SMT01}. In the model of \citet{ST99},
we have the cosmic average mass fraction of halos with mass higher
than $M$ given by
\be
F\SThalo(>M,z)=\int^{\infty}_{\sigma^2(R_c,z)}
f\ST(\delta_c,\sigma^2) d\sigma^2.
\ee

In a rare overdense region with volume much smaller than the universe,
the halo mass distribution significantly deviates from the cosmic mean
distribution shown by equations (\ref{eq:mf})-(\ref{eq:mfST}). Similar to
using the excursion set approach to get the extended Press-Schechter formula
\citep{Bondetal91,LC93}, we have the corresponding biased mass fraction of
$f\PS[\delta_c,\sigma^2(R_c,z)]$ in a region with comoving radius $R$ and mean
overdensity $\bar{\delta}_r$ at redshift $z$ given by
\be
&&f\bias\PS[\delta_c,\sigma^2(R_c,z);\bar{\delta}_r,R] \nonumber \\
&&=f\PS[\delta_c-\bar{\delta}_r,\sigma^2(R_c,z)-\sigma^2(R,z)],
\label{eq:fPSbias}
\ee
and the mass fraction of halos with mass higher than $M$ given by
\be
&&F\bias\PShalo(>M,z)=\nonumber \\
&&{\rm erfc} \left(\frac{\delta_c-\bar\delta_r}
{\sqrt{2[\sigma^2(R_c,z)-\sigma^2(R,z)]}} \right).
\label{eq:mfrac}
\ee
As shown in \citet{BL04}, the halo mass function in overdense regions
can be obtained by adjusting the Sheth-Tormen formula 
with a relative correction based on the extended Press-Schechter
model \citep{Bondetal91,LC93}, which can match cosmological
simulation results well. Similar to this hybrid model, the biased halo mass 
function in the regions with mean overdensity $\bar{\delta}_r$ can 
be obtained by\footnote{The biased halo mass function in a high-density region
may also be obtained by using the ellipsoidal collapse moving barrier model 
in \citet{ST02} instead of the hybrid model used here.}
\be
&&\left[\frac{dN}{dM}\right]\bias
=\frac{(1+\bar{\delta}_r) \bar{\rho}} {M} \nonumber \\
&& \times \left| \frac{d\sigma^2(R_c,z)}{dM}
\right| f\bias[\delta_c,\sigma^2(R_c,z);\bar{\delta}_r,R],
\label{eq:mfbiasst}
\ee
where
\be
&&f\bias[\delta_c,\sigma^2(R_c,z);\bar\delta_r,R]=
f\ST[\delta_c,\sigma^2(R_c,z)]\nonumber \\
&&\times \frac{f\bias\PS[\delta_c-\bar\delta_r,\sigma^2(R_c,z)-\sigma^2(R,z)]}
{f\PS[\delta_c,\sigma^2(R_c,z)]}.
\label{eq:mfbiasST}
\ee
In these overdense regions, the mass fraction of halos with mass higher than
$M$ is given by
\be
&&F\bias\halo(>M,z)=\nonumber \\
&&\int^{\infty}_{\sigma^2(R_c,z)}
f\bias(\delta_c,\sigma^2; \bar\delta_r,R)d\sigma^2.
\ee

In halos with sufficiently high masses, most of the gas in them is
collisionally ionized and some of the ionized gas may cool to form stars,
galaxies, and/or QSOs. Neutral hydrogen in the universe may then be
photoionized by photons from stars and QSOs.  We define a characteristic mass
scale $M\min$, the minimum mass of halos whose potentials are deep enough to
retain ionized gas in the photoionization equilibrium.  The temperature of
ionized gas in photoionization equilibrium is $\simeq 10^4$~K, and we have
$M\min=3.4 \times 10^7\msun [10/(1+z)]^{1.5}$ (see details for the
determination of this mass scale in Appendix B in \citealt{Benson01}).
We include both the gas outside halos with mass greater than $M\min$ and that inside these
halos but not collisionally ionized as IGM (the mass of baryons in galaxies is
negligible; see Fig.~\ref{fig:f1} below).  In these regions, the mass fraction
of hydrogen that is in IGM is given by
\be
&& x\HIGM(z)=\nonumber \\ &&1-
\int^{\infty}_{M\min}y\h\frac{M}{(1+\bar\delta_r) \bar{\rho}}
\left[\frac{dN}{dM}\right]\bias dM, \label{eq:fIGM}
\ee
where $y\h$ is the fraction of hydrogen that is collisionally ionized in a halo
with mass $M$ at the virial temperature $T\vir$, and the ratio of the baryonic
mass to the total mass in halos is assumed to be the same as the cosmic mean
$\Omega_{\rm b}/\Omega_{\rm m}$. 
The virial temperature is $T\vir=\frac{\mu m\h}{2k_B}\frac{GM}{r\vir}$, where
$\mu \sim 0.6$ is the mean molecular weight, $m\h$ is the proton mass, $k_B$ is
the Boltzmann constant, and $r\vir$ is the virial radius of the halo
\citep[see, e.g.,][]{BL01}.
The determination of the detailed distribution of collisionally ionized
hydrogen (see also eq.~\ref{eq:Clumphalo} below) would need to take into
account detailed gas dynamics inside the halos (e.g., photo-evaporation
processes, dynamics related to galactic disks), which is currently not well
understood. The $y\h$ used in this paper comes from \citet{SD93}, as commonly
adopted.

Our calculation shows that
in the rare overdense regions associated with the highest
redshift QSOs, $x\HIGM(r\Q,z\Q\simeq  6.2-6.4)\simeq 0.88-0.89$
while the cosmic mean $\bar{x}\HIGM(z\Q)$ is around $0.94$.

\subsection{The reionization due to stars} \label{sec:subreion}

Star formation in the formed halos is directly related to the cooling of their
virialized gas due to radiation by atomic, molecular, or metal lines.  Gas
cooling due to atomic lines, and thus star formation, can be very efficient in
halos with $T\vir>10^4$~K, as usually argued \citep[e.g.,][]{HS03}. However,
atomic line cooling is not efficient in halos with $T\vir<10^4$~K (the
so-called minihalos). %,
Molecular hydrogen (H$_2$) line cooling may be important for the very first
(Population III) star formation in those minihalos at very high redshift
(e.g., $z>10$; see \citealt{ABN02,BCL99}). These Population III stars may lead
to the reionization of the universe at very high redshift, but this may be
unimportant for the (second) reionization that finished at $z\sim 6$
\citep{Cen03a}. At lower redshift, external UV radiation may suppress the
molecular H$_2$ abundance in those minihalos and thus reduce the cooling due
to H$_2$ lines (e.g., \citealt{HRL96}). However, the precise cooling threshold and H$_2$
abundance in these regimes are not well known. Furthermore, star formation in
minihalos is also likely to be suppressed by the processes of
photoevaporation and supernova disruption.  Below we only consider
star formation in halos with $T\vir>10^4$~K and neglect star formation
in minihalos.  Detailed physical processes of star formation in halos with
$T\vir>10^4$~K are complicated and not fully understood.  Here we adopt two
simple models to estimate how many baryons form stars in these halos:

\begin{itemize}
\item
Model (a): The star formation efficiency (i.e., the mass fraction of baryons to
form stars) in halos is assumed to be independent of the halo mass, as adopted
in \citet{Cen03a}.  To satisfy the constraint that the reionization epoch ends
at $z\sim 6$, \citet{Cen03a} obtained a best-fit value of $0.1$ for the star
formation efficiency in halos with a virial temperature greater than $10^4$~K.  We adopt
this value and thus have the mass fraction of baryons that are in formed stars
in the rare overdense regions associated with the highest redshift QSOs,
$g\bias\star(z)\simeq 0.1 F\bias\halo(>M_4,z)$, where $M_4$ is
the mass of halos with virial temperature of $10^4$~K.

\item
Model (b): the star formation efficiency is assumed to depend on the halo mass.
After considering the feedback due to galactic winds in the star formation
process, \citet{HS03} argued that the normalized star formation rate (the mass
fraction of baryons to form stars per unit time) may maintain a roughly
constant level for halos with virial temperatures in the range
$10^4$-$10^{6.5}$~K but may rise to a level approximately 3 times higher for halos
with hotter virial temperatures. Based on this approximation, the star
formation rate in halos with temperature $T\vir$ is given by \citep[see eqs.
41, 43 and more details in][]{HS03}
\be
&&S(T\vir,z)\simeq \nonumber \\
&&\cases{
s_0 q(z) & {\rm for} $10^4{\rm K}<T\vir<10^{6.5}{\rm K}$, \cr
3s_0 q(z) & {\rm for} $T\vir> 10^{6.5}{\rm K}$, \cr
0 & {\rm otherwise},
}
\ee
where 
$q(z)=\left[ \frac{\tilde{\chi}\chi}{(\tilde{\chi}^m+\chi^m)^{1/m}}
\right]^{9/2\eta}$,
$\chi=[\Omega_{\rm m}(1+z)^3+\Omega_{\Lambda}]^{1/3}$, $m=6$, and $\eta=1.65$.
For the cosmological model adopted in this paper, we adjust $\tilde{\chi}$
from $4.6$ to $4.4$ and $s_0$ from $0.006h~\Gyr^{-1}$ to 
$0.008h~\Gyr^{-1}$ (see details for the dependence of $\tilde{\chi}$
and $s_0$ on cosmological parameters in \citealt{HS03}). 
Thus the mass fraction of baryons that are in formed stars in 
the rare overdense regions associated with the highest redshift QSOs 
is given by
\be
&& g\bias\star(z)\simeq \int^{\infty}_{z} s_0 q(z)\times \nonumber \\
&&[F\bias\halo(>M_4,z)+ 2F\bias\halo(>M_{6.5},z)]\frac{dt}{dz} dz,
\label{eq:starform}
\ee
while the corresponding cosmic mean is given by
\be
&& \bar{g}\star(z)\simeq \int^{\infty}_{z} s_0 q(z)\times \nonumber \\
&&[F\SThalo(>M_4,z)+ 2F\SThalo(>M_{6.5},z)]\frac{dt}{dz} dz,
\label{eq:starformbar}
\ee
where $M_{6.5}$ is the mass of a halo with virial temperature
$10^{6.5}$~K.
\end{itemize}

\begin{figure} 
\epsscale{1.2} 
\plotone{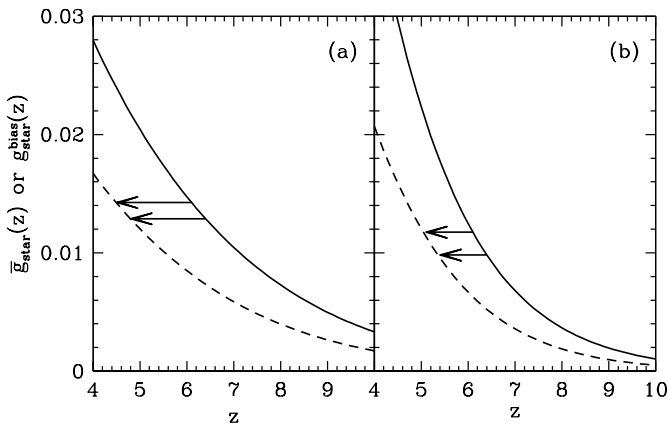} 
\caption{Mass fraction of baryons that are in formed stars as a function of 
redshift.
The solid line represents the fraction for spheres with proper radius
$r\Q=5\Mpc$ centering on the highest redshift QSOs [$g\bias\star(z)$;
eq.~\ref{eq:starform}], while the dashed line represents the cosmic mean
fraction [$\bar{g}\star(z)$; eq.~\ref{eq:starformbar}].  Panels (a) and (b)
show the results obtained from different star formation models (a) and (b)
(described in \S~\ref{sec:subreion}), respectively.  As seen from this figure,
although the mass fractions obtained from the two models are not exactly the
same, $g\bias\star(z\Q\sim 6.2-6.4)$ approaches the cosmic mean value
$\bar{g}\star(z)$ of a later time $z\sim 5$ in both models (indicated by the
arrows).  See details in \S~\ref{sec:subreion}.  } \label{fig:f1} \end{figure}

We plot the results of $g\bias\star(z)$ and $\bar{g}\star(z)$ in
Figure~\ref{fig:f1} for both models a and b. Their values at the QSO
redshift $z\Q\simeq 6.2-6.4$ are summarized in Table~\ref{tab:tab2}.  As seen
from Figure~\ref{fig:f1}, both models indicate that star formation in the
overdense regions ({\it solid line}) is significantly biased away from or higher than
the cosmic average at the same redshift ({\it dashed line}).  Although
the mass fractions of baryons in formed stars obtained from models a and b
are not exactly the same, the two models give a consistent result in that the
mass fraction of baryons in formed stars in the regions associated with the
highest redshift QSOs at $z\Q\sim 6.2-6.4$ is similar to the corresponding
cosmic mean values at a later time $z\sim 5$, i.e.,
$g\bias\star(z\Q)\simeq\bar{g}\star(z\sim5)$ (indicated by the arrows in
Fig.~\ref{fig:f1}; see also Tab.~\ref{tab:tab2}).  Since the number of ionizing
photons emitted from stars is proportional to the mass of baryons in formed
stars, the cumulative number of the ionizing photons produced in the stars in
the rare overdense regions associated with the highest redshift QSOs at
$z\Q\sim 6.2-6.4$ is thus similar to that of the corresponding cosmic mean at a
later time $z\sim 5$. However, their hydrogen ionization statuses may not be
naively similar, for example, because of differences in their proper densities
or in the process/time for hydrogen to dynamically respond to the ionization
(especially for the photo-evaporation process of minihalos).

We assume the initial mass function for star formation in the surrounding 
regions of the highest redshift QSOs is similar to that measured in the
nearby universe \citep{Scalo} and thus that roughly $4000$
ionizing photons are produced per baryon in the formed stars
\citep[e.g.,][]{BL01}. The rate of ejection of the ionizing photons
into the IGM can be given by 
\be
\dot{N}\bias\phsstar \simeq
\frac{4\pi}{3}R^3 (1+ \bar{\delta}_{r}) N_{\rm b}
\frac{dg\bias\star(z)} {dt} \times 4000f_{\rm esc},
\label{eq:nemit}
\ee
where $N_{\rm b}$ is the baryon comoving number density and $f_{\rm esc}$ is
the fraction of the ionizing photons escaping into the IGM and is roughly in
the range $0.1$-$0.2$ \citep[e.g.,][]{Cen03a}.
In spheres with proper radius
$r\Q=5\Mpc$ associated with the highest redshift QSOs, we obtain 
$\dot{N}\bias\phsstar(z\Q)$ for both models a and b and list them
in Table~\ref{tab:tab2}. For comparison, the cosmic mean
injection rate of ionizing photons due to stars at $z\sim 5$ in the same
comoving volume, $\bar{\dot{N}}\phsstar \simeq \frac{4\pi}{3}R^3 N_{\rm b}
\frac{d\bar{g}\star(z)} {dt} \times 4000f_{\rm esc}$, is also listed in
Table~\ref{tab:tab2}, which is roughly half of $\dot{N}\bias\phsstar(z\Q)$.
Assuming that the background ionizing energy intensity $J_{\nu}$
($\nu$: photon frequency) is mainly contributed by stars,
we estimate that an ionizing photon emission rate of $\bar{\dot{N}}\phsstar$ 
leads to the background intensity $J_{-21}$ at the Lyman limit in units of
$10^{-21}$ergs~cm$^{-2}$~s$^{-1}$~Hz$^{-1}$~sr$^{-1}$ of about
$0.06$-$0.13$ for model a and $0.1$-$0.2$ for model b
(using eq.~29 in \citealt{MPR99}).
These numbers are roughly consistent with those determined from observations at
$z\sim 5$ \citep[see Fig.~2 in][]{Fanetal02}, which suggests that the star
formation models used in this paper are roughly consistent with
observations. In the rare overdense regions around the highest redshift
QSOs, according to equation (\ref{eq:nemit}) the cumulative number of the 
ionizing photons emitted from stars before $t(z_Q)$ is 
about $5$-$10$ per hydrogen atom in the IGM within the Str\"omgren sphere,
which is substantially higher than that from the central QSO ($\la 1-2$;
see also discussion in \S~\ref{sec:model}).

As shown above, a large number of stars have already formed in the rare
high-density regions before $z\sim 6$, and in most of the time before the
cosmic time $t(z\Q)$, stars are the main sources
of the ionizing photons contributing to the re-ionization of neutral
hydrogen in the IGM (as mentioned in the introduction and as will also be 
discussed in \S~\ref{sec:model}).
The reionization starts in low-density regions (or voids) and then gradually
penetrates deeper
into high-density regions (see discussions in \citealt{Gnedin00,MHR00}). 
Thus, before the reionization by the QSO photons becomes effective,
the IGM in low-density regions surrounding those stars (or galactic
units) has already been highly ionized, while the
IGM in high-density regions may still be significantly neutral since the
ionizing photons from stars are still insufficient to ionize all the IGM. The
division from the highly ionized regions to the significantly neutral regions
 may be characterized by
a critical overdensity $\Delta\crit$, where the overdensity $\Delta$ is defined
through $\Delta\equiv\rho_b/\bar{\rho}_b$, $\rho_b$ is the comoving baryon mass
density at a given space position, and $\bar{\rho}_b$ is the cosmic comoving
mean mass density of baryons (note the difference from the definition of the
overdensity $\bar{\delta}_{r}$, which is averaged over a proper scale of $r$).
After the nuclear activity of the QSO turns on and the emission of its
ionizing photons becomes significant, QSO photons will ionize
some remaining neutral hydrogen (in high-density regions) not ionized by stars.
Compared to ionizing photons from stars, the photons from QSOs are much harder
and can penetrate into high-density regions more deeply. 
In addition, QSOs are
rare point sources located in high-density regions, while stars are embedded in
relatively numerous halos and act like diffuse sources.  A significant
fraction of the ionizing photons from stars may leak out of the regions (with
$r\Q=5\Mpc$) considered in this paper, while ionizing photons from QSOs do not,
but are absorbed. 

Here we give a rough estimate of
the ionization status due to stars in the overdense regions associated with the
highest redshift QSOs, such as the critical value $\Delta\crit$ and the
distribution of $\Delta$:
\begin{itemize}
\item 
We assume that the IGM is in the local photo-ionization-recombination
equilibrium, as commonly adopted \citep[see, e.g.,][]{Fanetal02}; that is, the
photoionization rate $n\hi \Gamma$ is balanced by the recombination rate
$n\hii n_{\rm e} \alpha_{\rm B}(T)$, where $n\hi$, $n\hii$, and $n_{\rm e}$ are
the proper number densities of neutral hydrogen, ionized hydrogen and electrons
in the IGM, respectively, $\Gamma$ is the number of photoionizations per H atom
per unit time and is roughly proportional to the ionizing photon emission rate,
and $\alpha_{\rm B}(T)$ is the recombination coefficient at temperature $T$
\citep{Abeletal97}. Note that $\dot{N}\bias\phsstar(z\Q\simeq 6.2-6.4) \simeq
2\bar{\dot{N}}\phsstar(z\sim5)$ and that the mean hydrogen density in the rare
overdense regions associated with the highest redshift QSOs at $z\Q$ is higher
than the cosmic mean at $z\sim 5$ by the same factor of $2$.  Thus, compared to
the photo-ionization equilibrium at $z\sim 5$, given the same ionization ratio
of $n\hii/n\hi$, the effect of a higher ionizing photon emission rate at $z\sim
6.2-6.4$ is compensated by the effect of a higher hydrogen density in the rare
overdense regions. Hence, we set the critical overdensity $\Delta\crit$ within
the rare overdense regions associated with the highest redshift QSOs at
$z\Q\sim 6.2-6.4$ to be similar to that in the cosmic average regions at $z\sim
5$.
The cosmic average ionization state of the IGM at redshift $z\sim 5$ has
been measured by \citet{Fanetal02} through the high-redshift QSOs around that
redshift and $\Delta\crit \simeq 10-20$ (see Fig.~8 in \citealt{Fanetal02}).
The simple arguments above have circumvented a more sophisticated and lengthy
way to obtain $\Delta\crit$, i.e., using $\dot{N}\bias\phsstar(z\Q)$ and
repeating some detailed formalisms used in \citet{MHR00}, which gives an even
larger $\Delta\crit$. Hereafter, we simply adopt $\Delta\crit \simeq 10-20$.
At least the regions with overdensity below this value can be highly
ionized by stars. This value is also compatible with the constraint
on the reionization process from the Str\"omgren sphere in \S~\ref{sec:Rsph}.
\item Since the mass fraction of halos with mass greater than $M\min$ within the rare
overdense regions associated with the highest redshift QSOs is similar to the
cosmic mean fraction at a later time $z\sim 4.5-4.8$ [see Fig.~\ref{fig:f1}a,
where $F\halo(>M_4,z)\simeq 10 g\star(z)$, or see Tab.~\ref{tab:tab2}],
we assume that the volume-weighted mass density distribution of baryons outside of
halos in these rare overdense regions is analogous to (or at least does not
significantly deviate from) the cosmic mean distribution at a later time
$z\sim 4.5-4.8$. Numerical simulations have shown that the cosmic mean
volume-weighted density distribution function $P(\Delta)$, defined so that
$P(\Delta)d\Delta$ represents the fraction of the volume occupied by regions
with overdensity in the range $\Delta\rightarrow\Delta+d\Delta$, is described
by \citep{MHR00} \be
P(\Delta)=A\exp\left[-\frac{(\Delta^{-2/3}-C_0)^2}{2(\delta_0/3)^2}\right]
\Delta^{-\beta}, \label{eq:probden} \ee where $A$ and $C_0$ are set by
normalizing $\int P(\Delta)d\Delta=1$ and $\int \Delta P(\Delta)d\Delta=1$, and
$\delta_0$ and $\beta$ are given by simulations, for example, $\delta_0=3$ and
$\beta=2.41$ at $z\sim 5$, and $\delta_0=3.6$ and $\beta=2.25$ at $z\sim 4$
\citep{Chiu03}.  The $P(\Delta)$ given by simulations may be biased away from
the reality because of the small size of the box adopted in numerical simulations
\citep{BL04}, but the bias is not much for the high redshift $z\Q\sim 6.2-6.4$
considered in this paper. 
\end{itemize}
Using the distribution $P(\Delta)$ given by numerical simulations at both
$z\sim 4$ and $z\sim 5$ (eq.~\ref{eq:probden}), we obtain that the mass
fraction of baryons or hydrogen in the regions with overdensity $\Delta \la
\Delta\crit \sim 10-20$ is about $x\h(\Delta\la\Delta\crit)=\int^{\sim
10-20}_0 \Delta P(\Delta)d\Delta \sim 0.7-0.8$.  Considering that the mass
fraction of hydrogen in IGM is $\sim 0.9$ in the regions associated with the
highest redshift QSOs as obtained in \S~\ref{sec:subhaloes} and that not only
the IGM with low densities ($\Delta\la\Delta\crit\sim 10-20$) but also part of
the IGM with high density ($\Delta\ga\Delta\crit$) may also be ionized by
photons from stars, the neutral hydrogen fraction is $x\hi\simeq
x\HIGM-x\h(\Delta\la\Delta\crit)\la 0.1-0.2$ in those rare overdense regions
before the reionization due to the QSO becomes effective.

Not only the models of structure and star formation above suggest that a
significant number of stars have formed in the early universe to contribute to
the reionization, there also exists emerging observational evidence
suggesting rapid star formation [with a rate of $\sim 3000\msun \yr^{-1}$,
which corresponds to an ionizing photon emission rate of 
$(0.5-0.9)\times 10^{56}\ps$ if the photon escaping fraction
$f_{\rm esc}\sim 0.1-0.2$] in the host galaxies
of the highest redshift QSOs. For example, (1) the high infrared
luminosity observed in one of the highest redshift QSOs, SDSS J1148+5251, suggests
that a large amount of dust has already formed \citep{Bertoldi03}, and (2) the
optical study of SDSS J1030+0524 suggests the existence of super-solar
metalicities in the highest redshift QSOs \citep{Pentericci}, which are also
expected to be the result of rapid star formation \citep{Walter03}.  The very
high star formation rate suggests that large elliptical galaxies or the host
galaxies of those QSOs with masses of $10^{11}-10^{12}\msun$ formed on a
dynamical timescale of a few times $10^8\yr$, which is roughly consistent with
current
scenario of the simultaneous formation of both QSOs/massive black holes and
elliptical galaxies.  If the mass of a massive black hole in those QSOs is around
a few $10^9\msun$, as estimated by \citet{WMJ03}, the mass ratio of black holes to
their host galaxies is $\sim 10^{-2}-10^{-3}$, which is close to the
ratio found in nearby massive black holes and galaxies
\citep[e.g.,][]{Magorrian,KG01}.

\subsection{The clumping factor of the hydrogen ionized by QSO photons}\label{sec:clump}

As discussed above, inhomogeneity or clumpiness of hydrogen may have a
significant effect on the reionization process.  We define a clumping factor
$C\hii\equiv \langle (n\hiiQ)^2\rangle/\langle n\h\rangle^2$ to describe the
small-scale clumpiness of the hydrogen ionized by QSO photons (not
collisionally ionized in halos or ionized by stars), where $n\hiiQ$ is the
proper number density of the hydrogen ionized by QSO
photons, and ``$\langle\cdot\cdot\cdot\rangle$'' represents the average over
the volume of the considered region.  
Note that our definition of the clumping factor here is a little different
from that conventionally defined as $\langle (n\h)^2\rangle/\langle 
n\h\rangle^2$, since the effects of stars and QSOs on the
reionization are isolated in this paper (see also eq.~\ref{eq:stromevol}).
In the regions with proper radius
${r\hii}$ and overdensity $\bar{\delta}_{r\hii}$ at redshift $z$, we have
$\langle n\h\rangle(z)=(1+\bar{\delta}_{r\hii})(1+z)^{3}\bar{N}\h$ and
$\bar{N}\h\simeq (1-3Y_{\rm He}/4)\bar\rho_{\rm b}/m\h$, where helium is
assumed to be singly ionized and its abundance $Y_{\rm He}=0.24$.  To estimate
$C\hii$, we divide the IGM in these regions into three components and then
consider the contribution by each one: (i) Gas in subregions with overdensity
$\Delta\la\Delta\crit\sim 10-20$, which has already been fully ionized by
photons emitted from stars, as argued in \S~\ref{sec:subreion}. This component
is not responsible for the absorption of the QSO ionizing photons, and thus
does not contribute to the clumping factor defined above. (ii) The remaining
neutral gas in halos with mass greater than $M\min$ (see the definition of $M\min$ in
\S~\ref{sec:subhaloes}), which has not yet been collisionally ionized and
cannot be photoevaporated out of the halo gravitational potential well. (iii)
Gas in the subregions with overdensity $\Delta\ga 10-20$ but outside of halos
with mass greater than $M\min$, which has not yet been fully ionized by photons from stars.
Below we estimate the contribution by components ii and iii.

For component ii, it is plausible to assume that the remaining neutral gas
only in the outer layer of these halos is ionized by the photons emitted from
stars, and most of the remaining neutral gas in component ii will be ionized
by QSO photons.  The average contribution of component ii
to $C\hii$ can be estimated by
\be
&&C^{\rm ii}\hii \simeq \frac{f_{\rm int} \Delta\vir}{
(1+\bar{\delta}_{\rm r\hii})^2}\times \nonumber \\
&&\int^{\infty}_{M\min}(1-y\h)^2\frac{M}{\bar\rho}
\left[\frac{dN}{dM}\right]\bias dM,
\label{eq:Clumphalo}
\ee
where $\Delta\vir\simeq 178$ is the virial overdensity and $f_{\rm int}$ is a
weight parameter for the distribution of gas density in halos and $f_{\rm
int}\simeq 3.14 $ if the halo has a nonsingular isothermal density profile
with a core radius of $0.1r\vir$ (see details in \S~4 and Appendix~B in
\citealt{Benson01}).  For spheres with proper radius $r\Q=5\Mpc$ centering on
the highest redshift QSOs at $z\Q\simeq 6.2-6.4$, we have the mean overdensity
$\bar{\delta}_{r\hii}=\bar\delta\rQ\simeq 0.25$ (see
\S~\ref{sec:overden}). Using equations (\ref{eq:mfbiasst}) and
(\ref{eq:Clumphalo}),  we find that
\be
C^{\rm ii}\hii \simeq 15-16.
\ee
The major contributors to $C^{\rm ii}\hii$ are halos with mass in the
range $M\min$--$3M\min$ and correspond to at most a mass fraction of
\be
&&\simeq \frac{1}{
(1+\bar{\delta}_{\rm r\hii})}\times \nonumber \\
&&\int^{\infty}_{M\min}(1-y\h)\frac{M}{\bar\rho}
\left[\frac{dN}{dM}\right]\bias dM\sim 2\%,
\label{eq:xhhalo}
\ee
of the total baryon. 
Our calculation also shows that the number of halos for ionizing photons
emitted from the central QSO at a random direction to pass through within the
scale of $r\Q=5\Mpc$ is on average one, which
is not too small to apply the average clumping factor at a random line of sight
obtained by equation (\ref{eq:Clumphalo}) to the particular observed QSOs.
However, for QSOs with lower luminosity (e.g., with ionizing photon emission
rate of less than $10^{56}~{\rm s}^{-1}$ at $z\Q\sim 6.2-6.4$), we find that the ionizing
photons from the QSO at a significant fraction of its lines of sight may not
encounter any halo with mass greater than $M\min$ within the Str\"{o}mgren sphere (but the
photons could be absorbed by component iii), and thus the value of $C^{\rm
ii}\hii$ will be an overestimate for the particular observed QSOs if they are
just located on these lines of sight (see discussions
in \S~\ref{sec:discussion}).

Compared to component ii, the contribution of component iii to $C\hii$ is
not easy to obtain accurately since the distribution of the IGM outside of
halos and the effects of the photoevaporation process on minihalos within the
rare overdense regions around the highest redshift QSOs are less well
understood.  To get the contribution of component iii, we consider the
regions with overdensity $70>\Delta\gtrsim 10-20$ (the upper limit 70 is set by
the local overdensity at the virial radius of a halo with a nonsingular
isothermal density profile and core radius $0.1r\vir$). Using the distribution
given by equation (\ref{eq:probden}) at $z\sim 4-5$, the clumping factor due to
gas in regions with overdensity between $\sim 10-20$ and $70$ is \be \sim
\frac{1}{(1+\bar{\delta}_{r\hii})}\int^{70}_{\sim10-20}\Delta^2 P(\Delta)
d\Delta \sim 2-3.  \label{eq:C21} \ee On the one hand, this value can only
be taken as a lower limit since the numerical simulations used to obtain
equation (\ref{eq:probden}) do not resolve halos with mass $\la 10^6\msun$ and
the clumpiness in these smaller scales may also contribute to $C\hii$. On
other hand, the clumping factor estimated above may be not significantly
smaller than the reality, since most of the gas in halos with mass $\la
10^6\msun$, especially in small halos, would be easily photoevaporated
\citep{BL99,SIR04}, and part of the gas in regions with overdensity between
$\sim 10-20$ and $70$ may actually be ionized by photons from stars. 

Note that the contribution from equation (\ref{eq:C21}) does not include the
contribution due to minihalos with mass between $M_{\rm res}$ and $M\min$,
which are also parts of component iii. 
With a simple model of the photoevaporation process,
\citet{HAM01} argued that these
minihalos are potentially important sinks of ionizing photons.  Using
numerical simulations of gas dynamics and radiation transfer in minihalos,
\citet{SIR04} argued that at $z=9$ it may take only about $(1-1.5)\times
10^8\yr$ to photoevaporate these minihalos by a source at a distance of
$1\Mpc$ away and emitting ionizing photons at a rate of $10^{56}\ps$ (or
equivalently, at a distance of $10\kpc$ away and with emission rate
$10^{52}\ps$), and thus the photoevaporation process may reduce the required
number of ionizing photons to reionize these minihalos.
Our calculations in \S~\ref{sec:subreion} show that the average ionizing
photons emitted from stars or galaxies at a distance of $1\Mpc$ from those
minihalos is about $(0.5-1.1) \times 10^{57}(1\Mpc/r)^2\ps\simeq
(0.2-0.4)\times 10^{56}\ps$ for model a and $(0.8-1.6) \times
10^{57}(1\Mpc/r)^2\ps\simeq (0.3-0.6) \times 10^{56}\ps$ for model b,
respectively. These rates are significantly lower than the required rate given
by \citet{SIR04} to evaporate the gas in minihalos within a period of
$(1-1.5)\times 10^8\yr$.  Therefore before the nuclear activity of the QSO
turns on, despite that much of the gas in minihalos may have already been ionized
and pushed out of the halo gravitational potential wells by the ionizing
photons from stars, there should still remain a considerable fraction of gas in
minihalos.  After the nuclear activity turns on, the remaining gas will be
ionized by QSO photons. As will be discussed in \S~\ref{sec:Rsph},
the time during which the QSO has radiated around
the observed luminosity $\dot{N}\phsQ^0$ before the cosmic time $t(z\Q)$
is probably at most a few times $10^7\yr$, which is 
shorter than the period $(1-1.5) \times 10^8\yr$ above;
thus not all of the gas is evaporated at $z\sim z\Q$.
Using the static approximation of the
photoevaporation process in minihalos,
\citet{BL99} show that
a significant fraction of the gas (about $20\%$-$40$\% for a typical QSO spectrum
at redshift $z\sim 8-20$) of big mini-halos ($\geq 10^6\msun$) is still
gravitationally bounded to the potential well after the reionization of the
universe is completed.  Based on the tendency of the fraction with redshift
shown in \citet{BL99}, this fraction could be larger at a lower redshift
$z\Q\sim 6.2-6.4$. Below we simply assume that half of the gas still remains in
minihalos and obtain the contribution of the minihalos to the clumping
factor as follows:
\be
0.5\times \frac{f_{\rm int} \Delta\vir}{(1+\bar{\delta}_{\rm r\hii})^2}
\int^{M\min}_{M_{\rm res}}\frac{M}{\bar\rho}
\left[\frac{dN}{dM}\right]\bias dM \nonumber \\
\simeq 17-19. 
\label{eq:fcmini}
\ee
Our calculation shows that within a proper scale of $r\Q=5\Mpc$, the number of
halos with mass between $M_{\rm res}$ and $M\min$ passed by QSO photons at a
random line of sight is on average $2-3$ before the photons are all absorbed,
so that the average clumping factor obtained above can be applied to 
the specific observed QSOs. 
Our calculation also shows that the fraction of hydrogen remaining in the
minihalos is
\be
0.5\times \frac{1}{(1+\bar{\delta}_{\rm r\hii})}
\int^{M\min}_{M_{\rm res}}\frac{M}{\bar\rho}
\left[\frac{dN}{dM}\right]\bias dM \simeq 3\%. 
\label{eq:xhmini}
\ee

Combining the contributions from components ii and iii, the clumping factor
in the spheres considered in this paper (with $r\Q=5\Mpc$ at $z\sim 6$) is
$$C\hii=C^{\rm ii}\hii+C^{\rm iii}\hii\sim (34-38).$$
To more accurately estimate the clumping factor would require very
high resolution numerical simulations with sufficiently large volume and
incorporating dynamical effects of photo-evaporation and radiation transfer and
simultaneous consideration of the ionizing effects from both stars and QSOs.  
The clumping factor $C\hii$ may slowly decrease with time elapse because of
the evolution of the gas density distribution
(e.g., due to photoevaporation of some minihalos).
Also, the distribution of (mini)halos within the Str\"omgren sphere
is likely to be nonuniformly clustered and could further introduce
fluctuations in $C\hii$ for different lines of sight.
In \S~\ref{sec:Rsph} we shall use not only the estimate above but also some
other different values of the clumping factor to illustrate the results.

\section{The evolution of the Str{\"o}mgren spheres around the highest redshift
QSOs} \label{sec:Rsph}

\subsection{Models}\label{sec:model}
After the nuclear activity of the highest redshift QSO turns on,
the neutral hydrogen remaining in high-density regions not ionized by stars will
be further ionized by QSO photons and thus the neutral hydrogen fraction in
the IGM surrounding the QSO will decrease.
A spherical ionization front centering on the QSO will appear,
separating the inside highly ionized HII region (due to QSO photons) and the
outside partly ionized region (due to stars). The expansion of the apparent
radius of the ionization front detected by observers
(or the Str\"omgren radius; see the apparent shape of the Str\"omgren sphere
in \citealt{Yu04}),
$r\hii$, can be described by the following equation for the 
number of hydrogen atoms ionized by the QSO in the Str\"omgren sphere:
\be
\frac{4\pi}{3}\frac{d(x\hi\langle n\h\rangle r^3\hii)}{d\tau}=
\dot{N}\phsQ(\tau) \nonumber \\
-\frac{4\pi}{3}\alpha_{\rm B}C\hii \langle n\h\rangle^2 r^3\hii,
\label{eq:stromevol}
\ee
where $\tau=t-t_i$ is the time that the QSO has passed at cosmic time $t$ since
its nuclear activity was triggered at cosmic time $t_i$,
$\dot{N}\phsQ(\tau)$ is the evolution of the QSO ionizing
photon emission rate, and $\alpha_{\rm B}=2.6\times 10^{-13}\cm^3\ps$ is the
hydrogen recombination coefficient to excited levels of hydrogen at temperature
$T=10^4$K (see Donahue \& Shull 1987; \citealt{SG87}; \citealt{CH00,MR00,Yu04}).
The terms on the right-hand side of equation (\ref{eq:stromevol}) account for
the ionization due to QSO ionizing photons and the recombination within the
Str\"omgren sphere, respectively.
The role of the recombination can be
characterized by a recombination timescale defined as follows:
\be
&&\tau\rec=x\hi(C\hii \langle n\h\rangle \alpha_{\rm B})^{-1} \nonumber\\
&&\simeq 4.8\times 10^7\yr\frac{x\hi}{(1+\bar{\delta}_{r\hii})} \left(
\frac{30}{C\hii} \right) \left(\frac{7.42}{1+z} \right )^3.
\label{eq:trec}
\ee
The recombination is negligible if $\tau\ll \tau\rec$, while important
if $\tau>\tau\rec$.

In equation (\ref{eq:stromevol}) the parameters
$x\hi$, $\langle n\h\rangle$, and $C\hii$ are the average values
within the dense region around a QSO with radius $r\hii$ at cosmic
time $t$, and their changes during the time interval $r\hii/c$ of the QSO photon
propagation in the Str\"omgren sphere are assumed to be negligible.
Their changes caused by the Hubble expansion are also negligible.
For simplicity, we shall also neglect their
dependence on $r\hii$ in the calculations below and set them to be the
values averaged within a fixed scale, e.g., $r\hii=r\Q$ at $z\Q$, as analyzed
in \S~\ref{sec:environ}. This simplification will not affect our conclusions.

For the evolution of the QSO ionizing photon emission rate $\dot N\phsQ$,
below we assume two simple models: in model i
$\dot{N}\phsQ=\dot{N}^0\phsQ$ is constant, and in model ii $\dot{N}\phsQ$
increases with time exponentially, i.e.,
$\dot{N}\phsQ(\tau)=\dot{N}^0\phsQ\exp[(\tau-\tau_{\rm Q})/\tau_S]$
($0<\tau\leq \tau_{\rm Q}$), where $\tau\Q=t(z\Q)-t_i$ is the age of the
QSO at $z\Q$, $\tau_S\simeq 4.5\times 10^7\yr[\epsilon/0.1(1-\epsilon)]$
is the Salpeter timescale, and $\epsilon$ is the mass-to-energy
conversion efficiency and is $\sim 0.1$ (e.g., \citealt{GSM04,YT02,YL04a}).
The ionizing photon emission rate is proportional to the QSO
luminosity if their intrinsic continuum spectra do not differ significantly.
\citet{YL04a} show that the exponential increase of the QSO luminosity
with a characteristic Salpeter timescale (that is, QSOs radiate at a rate close
to their Eddington luminosity) appears to be generally consistent with the
expected demographic relations between the QSO luminosity function and massive
dormant black holes in nearby galaxies. It is likely that the luminosity
evolution of a QSO also experiences a declining-luminosity phase (e.g, with the
consumption of the accreting materials at the end of the nuclear activity), but
this phase generally does not dominate the main population of luminous QSOs and
is not considered in this paper, for simplicity \citep{YL04a}. Models i and ii
are almost the same if $\tau_{\rm Q}\ll \tau_S$. 

\begin{figure}
\epsscale{1.2}
%\epsscale{0.8}
\plotone{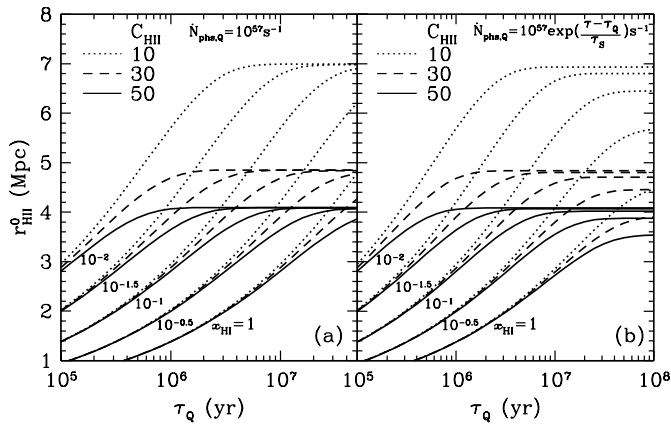}
\caption{
The expected apparent size of the Str\"omgren sphere $r\hii^0$ as a function of
the QSO age $\tau_{\rm Q}$. The QSOs are assumed to be at redshift $z\Q=6.3$.
Different line types represent different clumping factors used (dotted, dashed,
and solid lines for $C\hii=$10, 30, and 50, respectively).  For each line type,
we set the neutral hydrogen fraction $x\hi=10^{-2},
10^{-1.5},10^{-1},10^{-0.5}$, and 1 from top to bottom.  Panels (a) and (b) are
for different models of the evolution of the QSO ionizing photon emission rate
$\dot N\phsQ$ (models i and ii in \S~\ref{sec:model}, respectively).  As seen
from this figure, if $\tau_{\rm Q}$ is significantly short, $r\hii^0$ increases
with increasing $\tau\Q$ and decreasing $x\hi$ (see eq.~\ref{eq:trec}). If
$\tau_{\rm Q}$ is long enough (e.g., $>10^7\yr$), $r\hii^0$ approaches a
constant, independent of the detailed values of $\tau_{\rm Q}$.  The constant
is insensitive to the exact values of $x\hi$ and the evolution models of $\dot
N\phsQ$ used and is mainly determined by the clumping factor (the highest
redshift QSOs considered in this paper belongs to this case), unless $x\hi$ is
big enough (close to 1) in panel (b) (see eqs.~\ref{eq:strom2} and
\ref{eq:strom2s}).  See details in \S~\ref{sec:model}.
}
\label{fig:f2} 
\end{figure}

The highest redshift QSOs are also observationally
suggested to be accreting at a rate $\dot{M}_{\rm BH}$ close to the
Eddington limit \citep{WMJ03}. With this observational constraint,
in model i, the QSO age $\tau\Q$ should be not longer than
$M_{\rm BH}/\dot{M}_{\rm BH} \sim \tau_S$ (for this reason we extend
the $\tau\Q$-axis in Fig.~\ref{fig:f2}a at most to a few
times $10^7\yr$, rather than to $10^8\yr$ shown in Fig.~\ref{fig:f2}b),
and the cumulative ionizing photons emitted from the central QSO are at 
most $1$-$2$ per hydrogen atom in the IGM within their Str\"omgren spheres
(if $\epsilon\sim 0.1$), which is significantly less than those from stars
(see \S~\ref{sec:subreion}).
Although in model ii $\tau\Q$ may be much longer than $\tau_S$,
before $t(z\Q)$ most ionizing photons are emitted within the
period from $t(z\Q)-\tau_S$ to $t(z\Q)$. The time interval $\tau_S$
corresponds to a redshift interval of $\delta z\sim 0.3$ at $z\Q\sim 6.2-6.4$.
Cumulatively, most ionizing photons from stars before $t(z\Q)$ are emitted
before $t(z\Q+\delta z)$, and before $t(z\Q+\delta z)$ most ionizing photons
are not from QSOs but stars (see Fig.~\ref{fig:f1}), 
which helps to justify the isolation of the ionization effect due to stars and
that due to QSOs done in this paper (i.e., first consider stars and then QSOs).

Depending on whether the QSO shining time $\tau\Q$ is much shorter than the
recombination timescale $\tau\rec$ (and the Salpeter timescale $\tau_S$ for
model ii of $\dot N\phsQ$), equation (\ref{eq:stromevol}) and its solution can
be simplified into the following two cases:
\begin{itemize}
\item Case (1): if $\tau_{\rm Q}\ll\tau\rec$ (and $\tau_{\rm Q}\ll\tau_S$ for
model ii of $\dot N\phsQ$), the recombination term in equation
(\ref{eq:stromevol}) is negligible and the apparent size of the
highly ionized HII region at $\tau=\tau\Q$ 
is given by
\begin{eqnarray}
r\hii^0 & \simeq & \left(\frac{3\int_0^{\tau\Q}\dot{N}\phsQ(\tau) d\tau}
{4\pi x\hi\langle n\h\rangle} \right)^{1/3}, \\
& \simeq & \left( \frac{3\dot{N}\phsQ^0 \tau_{\rm Q}}
{4\pi x\hi\langle n\h\rangle} \right)^{1/3}.
\label{eq:strom1}
\end{eqnarray}
In this case, given $\dot{N}\phsQ^0$ and $\langle n_H\rangle$, the size of 
$r\hii^0\propto (\tau_{\rm Q}/x\hi)^{1/3}$
depends on the QSO age $\tau\Q$ as well as the neutral hydrogen fraction
$x\hi$.
There is little difference in the results from models i and ii.
\item Case (2): if $\tau\Q\gg\tau\rec$, 
the recombination within the highly ionized HII region is approximately
balanced by the emission of the ionizing photons from the QSO,
and we have
\begin{eqnarray}
r\hii^0 & \simeq & \left( \frac{3\dot{N}\phsQ^0\tau\rec}{4\pi x\hi\langle n\h\rangle}\right)^{1/3}, \nonumber \\
& = & \left( \frac{3\dot{N}\phsQ^0}{4\pi\alpha_{\rm B}
C\hii\langle n\h\rangle^2}\right)^{1/3}\equiv r_S
\label{eq:strom2}
\end{eqnarray}
for model i and 
\be
r\hii^0\simeq\left(\frac{\tau_S}{\tau\rec+\tau_S}\right)^{1/3}r_S
\label{eq:strom2s}
\ee
for model ii.
The definition of $r_S$ above is similar to the classical formula of the
Str\"omgren radius in stellar astronomy. In this case, given $\dot{N}\phsQ^0$
and $\langle n_H\rangle$, $r\hii^0$ does not depend on the exact value of the
QSO age $\tau_{\rm Q}$.  If $\tau_S\gg\tau\rec$ in model ii, there is little
difference in the results of the two models and $r\hii^0(\propto C\hii^{-1/3})$
increases with decreasing $C\hii$.  The dependence of $C\hii$ on $x\hi$ is
probably weak since most hydrogen atoms are actually located in
low-density regions, while $C\hii$ is mainly determined by the hydrogen in
high-density regions \citep[as illustrated in Fig.~2 in][]{MHR00}. If
$\tau_S\ll\tau\rec$ in model ii, $r\hii^0$ obtained from model ii is
smaller than that from model i, and $r\hii^0(\propto x\hi^{-1/3})$ depends
mainly on $x\hi$ in model ii instead of $C\hii$ in model i.  \end{itemize}

Below using the $\dot N\phsQ$ given by models i and ii, we solve equation
(\ref{eq:stromevol}) to illustrate the general properties of the evolution of
$r\hii$ by artificially setting different values for the clumping factor $C\hii$
and the neutral hydrogen fraction $x\hi$. The solutions of $r\hii^0$ are shown
as a function of the assumed QSO age $\tau_{\rm Q}$ in Figure~\ref{fig:f2}.
Panels (a) and (b) give the results for models i and ii of $\dot{N}\phsQ$,
respectively.  In both panels, the QSOs are assumed to be at $z\Q=6.3$ and have
$\dot{N}^0\phsQ=10^{57}{\rm erg~s^{-1}}$, and we set
$\bar{\delta}_{r\hii}\simeq 0.25$, $C\hii$=10, 30, and 50 ({\it
dotted, dashed, and solid lines, respectively}), and
$x\hi=10^{-2},10^{-1.5},10^{-1},10^{-0.5}$, and 1 (top to bottom for each
line type).  As seen from both panels, at the short-age end (e.g.,
$<10^7\yr$), $r\hii^0$ increases with increasing $\tau_{\rm Q}$ and decreasing
$x\hi$ as described by case 1 above in which the effect of recombination is
insignificant (see eq.~\ref{eq:strom1}); while at the long-age end,
with increasing QSO ages, $r\hii^0$ roughly approaches a constant and the
constant increases with decreasing $C\hii$ as described by case 2.  Given
$C\hii$, the value of the constant is generally independent of $x\hi$ but may
decrease with increasing $x\hi$ if $x\hi$ is sufficiently large (close to 1 so
that $\tau_S\la\tau\rec$) in Figure 2b for model ii (see eqs.~\ref{eq:strom2}
and \ref{eq:strom2s}).

\subsection{Comparison with observations}\label{sec:comparison}
For the highest redshift QSOs considered in this paper
($\bar{\delta}_{r\hii}\simeq \bar{\delta}\rQ \simeq 0.25$, $x\hi\la 0.1-0.2$,
$C\hii\sim 36$, $z\Q\sim$6.2--6.4; see details in \S~\ref{sec:subreion} and
\S~\ref{sec:clump}), the averaged recombination timescale of the hydrogen in
their surrounding regions is about $\tau\rec\la 5\times 10^6\yr$.  Note that
the average lifetime of the main population of QSOs (with comoving number
density peaked at $z\sim$2--3) has been determined to be $\tau_{\rm life}
\gtrsim 4\times 10^7\yr$ \citep[][see also Martini 2004 for a review of the QSO
lifetime]{YT02,YL04a}.  If the lifetimes of the highest redshift QSOs are also
$\gtrsim 4\times 10^7\yr$, the
probability that their ages are shorter than $\tau\rec$ is therefore only $\sim
\tau\rec/ \tau_{\rm life}\la 10\%$, and the ages of the majority of the
observed highest redshift QSOs are more likely to be around or longer than a
few times $10^7\yr$, which is significantly longer than
$\tau\rec$. 

Even without the constraints of the QSO lifetime from
other methods and an accurate estimate of $C\hii$, by establishing a
statistical method relating the distribution of the observed Str\"omgren radii
with the probability distribution of the ages of the observed QSOs (somewhat
similar to the manipulations in \citealt{YL04a,YL04b}), one can still check or
rule out the possibility of whether the lifetime of the highest redshift QSOs is
shorter than $\tau\rec$ or whether the solution of $r\hii^0$ belongs to case
1.  For example, the observational distribution of $(\langle
n\h\rangle/\dot{N}^0\phsQ)^{-1/3}r\hiiobs^0$ can be estimated from a sample of
QSOs having Gunn-Peterson troughs.  This distribution should be consistent with
a constant if most of the $r\hii^0$ solutions belong to case 2, and consistent with the
distribution of $\tau_{\rm Q}^{1/3}$ obtained by a random choice of $\tau\Q$
for each QSO if most of the $r\hii^0$ solutions belong to case 1 and $x\hi$ is irrelevant
to $\tau_{\rm Q}$ and does not differ much for different QSOs in the sample.
Our preliminary results of applying this statistical method to the three
highest redshift QSOs with measured Str\"omgren radii (in Tab.~\ref{tab:tab1})
have shown that the possibility of $\tau_{\rm life}<\tau\rec$ is less than
15\%.  For simplicity, the details of this method are not presented in this
paper and will be deferred to future work with an increasing number of the
detected highest redshift QSOs having Gunn-Peterson troughs. 

We conclude here
that for most highest redshift QSOs the solution of $r\hii^0$
belongs to case 2 and is insensitive to the exact values of the QSO age.
For the highest redshift QSOs, we also have $\tau\rec\ll\tau_S$ (in model ii),
and thus $r\hii^0$ is also insensitive to the neutral hydrogen fraction and the
detailed evolution models of $\dot N\phsQ$ (see eqs.~\ref{eq:strom2} and
\ref{eq:strom2s}). The $r\hii^0$ value is mainly determined
by the QSO ionizing photon emission rate $\dot{N}^0\phsQ$ and the clumping
factor $C\hii$. 
Below we simply show the result obtained from model i.

\begin{figure} \epsscale{1.0}
%\epsscale{0.8}
\plotone{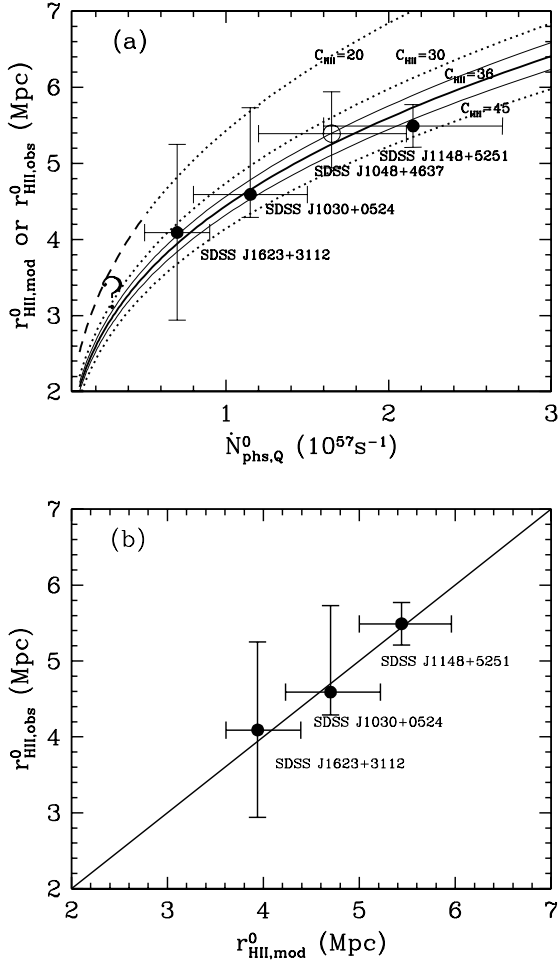} \caption{ Panel (a): The size of the highly ionized HII
regions $r\hii^0$ of the highest redshift QSOs as a function of the ionizing
photon emission rate $\dot{N}^0\phsQ$. The filled circles represent the
observational data of the three QSOs with measured $r\hiiobs^0$ given in the
literature (see Tab.~\ref{tab:tab1}).  The open circle gives our model results
of $r\hiimod^0$ for the QSO J1048+4637 obtained by solving equation
(\ref{eq:stromevol}) and setting $\dot{N}^0\phsQ$ to be the mean of its upper
and lower limits (listed in Table~\ref{tab:tab1}), and its estimated error bar
in $r\hiimod^0$ represents the range of the upper and lower limits of
$\dot{N}^0\phsQ$ (similarly for the model results of other sources shown in
panel b).  The dotted and solid curves also represent the model results.  The
dotted lines give the results for QSOs at $z\Q=6.3$, and the solid lines give
the results for QSOs at $z\Q=6.2,6.3,6.4$ from top to bottom, respectively.
The top and bottom dotted lines are obtained by setting $C\hii$=20, 30 and 45,
respectively; and the solid lines and the open circle are obtained by setting
$C\hii=36$.  Note that for low-luminosity QSOs (with $\dot{N}^0\phsQ$ smaller
than a few times $10^{56}~{\rm s}^{-1}$) located in less dense regions,
$C\hii=36$ could be an overestimate and the range of observed $r\hii^0$ could
be higher than those shown by the dotted and dashed lines and up to the dashed
line (obtained by $C\hii=20$; see the region labeled by ``?'').  Panel (b): the
model results of the apparent size of the Str\"omgren sphere $r\hiimod^0$
versus the size obtained from observations for the highest redshift QSOs
$r\hiiobs^0$ (solid circles).  The solid line is a reference line.  The
remarkable consistency of the model results with observations supports that the
observed Str\"omgren radii can be self-consistently explained by the dense
environment around the highest redshift QSOs, the corresponding biased
structure formation, star formation, and reionization processes analyzed in
this paper, and the QSO lifetime longer than a few times $10^7\yr$. If the
observational $r\hiiobs^0$ is systematically underestimated, then the realistic
$C\hii$ may be lower than $\sim 36$ and the photo-evaporation process in minihalos may be
stronger than that assumed in this paper. See
details in \S~\ref{sec:Rsph}. 
} \label{fig:f3} \end{figure}

We plot $r\hii^0$ as a function of the ionizing photon emission rate of QSOs
$\dot N\phsQ^0$ in Figure~\ref{fig:f3}a. The observationally determined
Str\"omgren radii of three QSOs and their ionizing photon emission rates (see
Tab.~1) are shown by filled circles.  The model results obtained by setting
$\bar{\delta}_{r\hii}\simeq \bar{\delta}\rQ \simeq 0.25$, $C\hii=36$, and
$x\hi\la 0.1-0.2$ are shown by the dotted and solid curves, and by an open
circle (in the range of $r\hii^0\sim4.9-5.9\Mpc$) for the source SDSS
J1048+4637, which does not have a measured Str\"omgren radius in the literature.
The different solid lines are for QSOs at different redshifts ($z\Q=6.2$, 6.3,
and 6.4;
top to bottom).  We also illustrate the effects of $C\hii$ by showing the
results obtained by three other values of $C\hii$=20, 30, and 45 in
Figure~\ref{fig:f3}a (top, middle, and bottom dotted lines, respectively; $z\Q=6.3$).
The relation between the model and observational results for the three QSOs are
more explicitly shown in Figure~\ref{fig:f3}b (see solid circles).  As seen
from Figure~\ref{fig:f3}, the model results are remarkably consistent with
observations, which supports that the lifetimes of the highest redshift QSOs are
around or longer than a few times $10^7\yr$, as is the lifetime of the main population of QSOs
\citep[e.g.,][]{YT02,YL04a}. The remarkable consistency also supports that the
observational sizes of the Str\"omgren spheres of the highest redshift QSOs can
be self-consistently explained by the fact that they are located in rare overdense
regions and that the neutral hydrogen fraction in these regions is in the range from
several percent to 10\%--20\% (before considering the reionization due to QSO
photons).  A much smaller neutral hydrogen fraction (e.g., $\ll 3\%$) will not
be consistent with the clumping factor used, since $C\hii\sim 36$ implies that a
significant fraction of the neutral hydrogen (see eq.~\ref{eq:xhmini}) remains in
(mini)halos (which was neither ionized by the ionizing photons from stars nor
photo-evaporated out of the halos) and contributes to $C\hii$ (see
eq.~\ref{eq:xhmini}).  Unless $C\hii$ is much smaller than the estimate in this
paper (e.g., because of stronger photo-evaporation of baryons in minihalos), a
much shorter QSO lifetime ($\ll 10^7\yr$) will not be consistent with the
observations of $r\hiiobs^0$ since $r\hii^0$ decreases with decreasing $\tau\Q$
at short-$\tau\Q$ ends (see Fig.~\ref{fig:f2}).  

The solution of $r\hii^0$ in model ii is lower than that in model i by at most
$1-[\tau_S/(\tau\rec+\tau_S)]^{1/3}\sim \tau\rec/(3\tau_S)$. An independent
constraint on the timescale $\tau_S$ may be obtained by comparing the result of
model ii with the observation, as shown in Figure~\ref{fig:f3}b, but this may
require precise measurements of the observational Str\"omgren radius
$r\hiiobs^0$.  The current observational error of $r\hiiobs^0$ ($\delta
r\hiiobs^0\sim$5\%--30\%; see Tab.~\ref{tab:tab1})
and the required consistency with the observation may
give the constraint to be $\tau_S/\tau\rec\ga 1/(3\delta r\hiiobs^0)\sim$1--7.

\subsection{Discussions}\label{sec:discussion}
The model results of $r\hiimod^0$ above are obtained by using the average
clumping factor (over all the lines of sight).  A significant contribution to
the clumping factor comes from the neutral hydrogen in halos. In sparse
regions, the ionizing photons may have been absorbed before they encounter any
halo with mass greater than $M\min$ at a significant fraction of the
lines of sight.  Thus the average contribution of component iii to $C\hii$
may be an overestimate when applying it to a specific line of sight (see
discussions in \S~\ref{sec:clump}).  Hence, the model results above are usually
appropriate to be applied to the observed high-luminosity QSOs located in dense
regions, but are underestimates for low-luminosity QSOs (e.g., ionizing photon
emission rate $\sim10^{56}\ps$ at $z\Q\sim$6.2--6.4) located in less dense
regions.  The expected range of $r\hii^0$ for low-luminosity QSOs can be
estimated in the following way. By starting with their number density $N_{\rm
Q}$ (obtained by extrapolating the luminosity function of the highest redshift
QSOs with a power-law index of $3.2$; see details in
\citealt{Fanetal03,Fanetal04}) at the faint end and doing similar quantitative
analysis for the environment around the low-luminosity QSOs as done for the
high-luminosity QSOs above, we find that the clumping factor $C\hii$ is around
$40$ for those lines of sight along which one halo with mass greater than $M\min$
can be encountered within the Str\"{o}mgren sphere,
and around $20$ for other lines of sight (e.g., 50\% of all the lines of sight)
along which no halo with mass greater than $M\min$ is encountered (the details
of the calculations are not given here for simplicity).  Therefore, at a random
line of sight, the size of the Str\"omgren spheres around the low-luminosity
QSOs are roughly in the range of the results obtained by using $C\hii\sim20$
(dashed line in Fig.~\ref{fig:f3}a) and $C\hii\sim 36$ (see the region labeled
by ``?'' between the dashed line and the solid lines in Fig.~\ref{fig:f3}a).
We believe that comparison of the range with future observations of the
Str\"omgren radii of low-luminosity QSOs would further give some tests or
constraints on the analysis or assumptions made in this paper, for example, on
whether the major part of the clumping factor is due to the remaining neutral
hydrogen in halos.

In addition, we note that a higher neutral hydrogen fraction ($x\hi>0.3$ if
$\tau_{\rm Q}>10^7$~yr) is obtained in \citet{WL04}, but it is obtained by
using a much smaller clumping factor and without
considering the detailed effects of the structure and star formation in the
dense environment surrounding the highest redshift QSOs.  Also, as mentioned in
\S~\ref{sec:data} the Str\"{o}mgren radius of SDSS 1030+0524 estimated by
\citet{MH04} through the Ly$\beta$ trough in the QSO spectra ($\sim 6\Mpc$) is
larger than the value ($\sim 4.6\Mpc$) listed in Table \ref{tab:tab1}.  
A possibly larger observational Str\"{o}mgren radius may suggest that the
realistic clumping factor for this QSO is smaller than the estimate in this
paper (e.g., $C\hii$ may be as low as $\sim 20$; see Fig.~\ref{fig:f2}).  A
smaller $C\hii$ may be caused by a stronger photoevaporation process in
minihalos than that assumed in this paper.  If more minihalos were
photoevaporated, besides $C\hii$, the neutral hydrogen fraction to be ionized
by the central QSO, $x\hi$, may also decrease. In short, most of the analysis
in this paper will not be qualitatively affected even if the Str\"{o}mgren 
radii estimated from current observations deviate slightly from 
reality.

As indicated in Figure~\ref{fig:f1}, 
in the rare overdense regions around the highest redshift QSOs, the ionizing
photon production per hydrogen atom in the IGM at $z\sim 7$
is already comparable to or higher than that of the cosmic average
at $z\sim 6$ (when the reionization is complete). However, the reionization in 
the rare overdense regions around the highest redshift QSOs is
not complete at a redshift significantly higher than $z\Q\sim 6.2-6.4$ since
neutral hydrogen must remain in the high-density regions to be ionized by
QSO photons to satisfy the observational constraint from
the Str\"omgren radius analyzed in this paper.
The incomplete reionization in the overdense regions is partly
caused by the fact that they have more high-density gas and that the gas is easier to
recombine. This environmental effect has also been revealed in the
numerical simulation by \citet{Ciardi03}, which shows that
the reionization process in overdense regions may be complete
at a time later than that in cosmic average regions,
although the ionizing photon production per hydrogen atom in the IGM
is higher in overdense environments than in cosmic average regions.
According to this environmental effect revealed by \citet{Ciardi03}, the
cosmic average neutral hydrogen fraction at $z\sim 6.2-6.4$ should be not
higher than 10\%--20\% (the upper limit of $x\hi$ surrounding the highest 
redshift QSOs). The observational flux upper limit of the Gunn-Peterson 
troughs has constrained the lower limit of the neutral hydrogen fraction 
(in mass) of the universe at $z\sim 6.2-6.4$ to be 1\% \citep{Fanetal02}. 
Therefore, the cosmic average neutral hydrogen fraction at $z\sim 6.2-6.4$
may be only a few percent. This low neutral hydrogen fraction alleviates the
apparent conflict in reionization between the constraint from 
the highest redshift QSOs (e.g., \citealt{WL04}) and that from the
polarization spectrum of the cosmic microwave background 
\citep[i.e., an early reionization at $z\sim 15$ or so;][]{Kogut03,Spergel03}.

\section{Conclusions}\label{sec:con}

In this paper we have investigated the dense environment, the reionization
process, and the evolution of the Str\"omgren sphere around the highest
redshift QSOs having Gunn-Peterson troughs ($z>6.1$), and we have provided 
constraints on the intrinsic properties of QSOs and the reionization history of
the universe by comparing the observed Str\"omgren radii with model results.

We have shown that the structure formation and consequently the gas distribution
and star formation in the overdense regions around the highest redshift QSOs are
biased from the cosmic average.
Before the nuclear activity of the QSO turns on and its ionizing photon emission
rate is high enough, the reionization in the overdense region surrounding the
QSO is mainly contributed by stars, and starts from relatively low density
subregions. Using some simple models of
star formation and some analysis on the photoionization process inside and
outside of halos (including minihalos), we have argued that a significant
fraction of hydrogen in the Str\"omgren sphere around the QSOs is 
ionized by the ionizing photons from stars, and only about several percent to at most
10\%--20\% of hydrogen is left (e.g., in minihalos, halos, or
high-density subregions) to be ionized by the QSO photons. 
The cosmic average neutral hydrogen fraction at $z\sim 6.2-6.4$ is also be
smaller than the upper limit of 10\%--20\% obtained for the overdense
regions and may be only a few percent,
since overdense regions contain more high-density gas and are more difficult to
reionize \citep{Ciardi03}.
We have analyzed the clumping
property of hydrogen ionized by QSOs and studied the evolution of the apparent
size of the Str\"omgren sphere.  We have found that if the QSO lifetime is about 
or longer
than a few times $10^7$ yr, as is the lifetime of the main population of QSOs (with comoving number
density peaked at $z\sim$2--3), the expected Str\"omgren radii from our models
are very consistent with observations. With such a QSO lifetime, the ages of most
of observed QSOs are long enough that the QSO photon emission is balanced by
the recombination of the hydrogen ionized by QSOs in their Str\"omgren spheres,
and the expected Str\"omgren radii from the balance are independent of the
detailed values of the QSO ages.  We also point out a statistical method
involving a larger sample of highest redshift QSOs having Gunn-Peterson troughs
in future observations, which may potentially check or rule out the possibility
that the QSOs have a shorter lifetime (e.g., $<10^7\yr$) even without an
accurate estimate of the hydrogen clumping property.

We thank the referee for useful comments.
Q.Y. acknowledges support provided by NASA through Hubble Fellowship grant
\#HF-01169.01-A awarded by the Space Telescope Science Institute, which is
operated by the Association of Universities for Research in Astronomy, Inc.,
for NASA, under contract NAS 5-26555. Q.Y. acknowledges the hospitality of
the Aspen Center for Physics, where part of this work was completed.

\begin{deluxetable}{ccccccr}
\tablecolumns{7}
\tablewidth{0pt}
\tablecaption{Sample of QSOs that have the Gunn-Peterson trough}
\tablehead{
\colhead{Name} & \colhead{$z\Q$} & \colhead{$M_{1450}$} & 
\colhead{$\dot{N}^0\phsQ$}  &
\colhead{$z\hiiobs$} &
\colhead{$r\hiiobs^0$} & 
\colhead{$r\hiimod^0$}\\
\colhead{}    &\colhead{}     &\colhead{}  & \colhead{($10^{57}\ps$)}  &
\colhead{} & \colhead{(Mpc)} & \colhead{(Mpc)} }\\
\startdata
SDSS J103027.10+052455.0 & 6.28$^{+0.02}_{-0.005}$ & -27.15 & $1.0^{+0.2}_{-0.2}$-$1.2^{+0.3}_{-0.2}$ & 6.20-6.28 & $4.6^{+1.1}_{-0.3}$
& 4.2-5.2\\
SDSS J104845.05+463718.3 & 6.19$\pm$0.005 & -27.55 & $1.5^{+0.3}_{-0.3}$-$1.7^{+0.4}_{-0.3}$ & \nodata-6.19 & \nodata & 4.9-5.9\\
SDSS J114816.64+525150.3 & 6.42$\pm$0.005 & -27.82 & $1.9^{+0.4}_{-0.3}$-$2.2^{+0.5}_{-0.4}$ & 6.32-6.42 & 5.5$\pm$0.3 & 5.0-6.0\\
SDSS J162331.81+311200.5 & 6.22$\pm$0.02 & -26.67 & $0.6^{+0.2}_{-0.1}$-$0.7^{+0.2}_{-0.1}$ & 6.15-6.22 & 4.0$\pm$1.1 & 3.6-4.4\\
\enddata
\tablecomments{The $z\Q$, $M_{1450}$, and $\dot{N}^0\phsQ$ are the redshift,
absolute magnitude at $1450\AA$, and ionizing photon emission rate of the QSOs.
The $z\hiiobs$ gives the range between the QSO redshift and the redshift of
the onset of the Ly$\alpha$ Gunn-Peterson trough. The
$r\hiiobs^0$ gives the size of the Str\"omgren sphere that $z\hiiobs$
corresponds to.  For details about the
observationally determined data and their errors, see \S~\ref{sec:data}.  
The $r\hiimod^0$ is the Str\"omgren radius expected from the models
in this paper (see also Fig.~\ref{fig:f3}b).
}

\label{tab:tab1}
\end{deluxetable}

\begin{deluxetable}{cccccr}
\tablecolumns{6}
\tablewidth{0pt}
\tablecaption{Mass fraction of baryons in formed stars and ionizing photon
emission rates from stars}
\tablehead{
\colhead{Model} & \colhead{$g\bias\star(z\Q)$} &
\colhead{$\bar{g}\star(z\Q)$} &
\colhead{$z$}  &
\colhead{$\dot{N}\bias\phsstar(z\Q)$} &
\colhead{$\bar{\dot{N}}\phsstar(z\sim 5)$} \\
\colhead{}    &\colhead{($\times 10^{-2}$)}    
&\colhead{($\times10^{-2}$)}  & \colhead{$\bar{g}\star(z)=g\bias\star(z\Q)$}  &
\colhead{($10^{57}\ps$)} & \colhead{($10^{57}\ps$)} }\\
\startdata
(a) & 1.3-1.4 & 0.8 & 4.5-4.8 & 0.5-1.1 & 0.3-0.6\\
(b) & 1.0-1.2 & 0.6 & 5.0-5.3 & 0.8-1.6 & 0.4-0.8\\
\enddata
\tablecomments{Models (a) and (b) are two different models for the star
formation efficiency (i.e., the mass fraction of baryons to form stars) in
halos (see details in \S~\ref{sec:subreion}). The $\bar{g}\star(z\Q)$ gives
the mass fraction of baryons that are in formed stars in the rare overdense
regions
with proper radius $r\Q=5\Mpc$ around the highest redshift QSOs, where the QSO
redshift $z\Q\simeq 6.2-6.4$; and the $\bar{g}\star(z\Q)$ gives the cosmic mean
fraction at the same redshift. The fourth column gives the redshift at which
$\bar{g}\star(z)=g\bias\star(z\Q)$ (indicated by the arrows in
Figure~\ref{fig:f1}.  The redshift is close to $z\sim 5$ for both models.  The
$\dot{N}\bias\phsstar(z\Q)$ gives the ionizing photon emission rates from stars
in the overdense regions associated with the highest redshift QSOs at redshift
$z\Q$; and the $\dot{N}\phsstar(z\sim 5)$ gives the cosmic mean rate at
$z\sim5$.  We have $\dot{N}\bias\phsstar(z\Q)\simeq
2\bar{\dot{N}}\phsstar(z\sim 5)$ for both models.  See also Fig.~\ref{fig:f1}
and \S~\ref{sec:subreion}.  } \label{tab:tab2} \end{deluxetable}

\end{document}